# Orbital hybridization-driven charge density wave transition in $CsV_3Sb_5$ Kagome superconductor

*Shulun Han[1,#], Chi Sin Tang[1,2,#], Linyang Li[3,#], Yi Liu[4,#], Huimin Liu[1], Jian Gou[5,6], Jing Wu[7], Difan Zhou[1], Ping Yang[2], Caozheng Diao[2], Jiacheng Ji[1], Jinke Bao[1], Lingfeng Zhang[1,\*], Mingwen Zhao[8], M. V. Milošević[9], Yanqun Guo[1], Lijun Tian[1], Mark B. H. Breese[2,5], Guanghan Cao[10], Chuanbing Cai[1], Andrew T. S. Wee[5,6,\*], Xinmao Yin[1,\*]*

[1]Shanghai Key Laboratory of High Temperature Superconductors, Shanghai Frontiers Science Center of Quantum and Superconducting Matter States, Physics Department, Shanghai University, Shanghai 200444, China

[2]The Singapore Synchrotron Light Source (SSLS), National University of Singapore, Singapore 117603

[3]School of Science, Hebei University of Technology, Tianjin 300401, China

[4]Department of Applied Physics, Zhejiang University of Technology, Hangzhou 310023, China.

[5]Department of Physics, Faculty of Science, National University of Singapore, Singapore 117542

[6]Centre for Advanced 2D Materials and Graphene Research, National University of Singapore, Singapore 117546

[7]Institute of Materials Research and Engineering, Agency for Science, Technology and Research (A*STAR), 2 Fusionopolis Way, Singapore, 138634 Singapore

[8]School of Physics, Shandong University, Jinan 250100, China

[9]Departement Fysica, Universiteit Antwerpen, Groenenborgerlaan 171, B-2020 Antwerpen, Belgium

[10]Department of Physics, Zhejiang University, Hangzhou 310027, China

*Correspondence to: yinxinmao@shu.edu.cn (X.Y.); lingfeng_zhang@shu.edu.cn (L.Z.); phyweets@nus.edu.sg (A.T.S.W.).

[#]These authors contributed equally to this work.

**Abstract:** Owing to its inherent non-trivial geometry, the unique structural motif of the recently discovered Kagome topological superconductor $AV_3Sb_5$ (A = K, Rb, Cs) is an ideal host of diverse topologically non-trivial phenomena, including giant anomalous Hall conductivity, topological charge order, charge density wave (CDW), and unconventional superconductivity. Despite possessing a normal-state CDW order in the form of topological chiral charge order and diverse superconducting gaps structures, it remains unclear how fundamental atomic-level properties and many-body effects including Fermi surface nesting, electron-phonon coupling, and orbital hybridization contribute to these symmetry-breaking phenomena. Here, we report the direct participation of the V3$d$—Sb5$p$ orbital hybridization in mediating the CDW phase transition in $CsV_3Sb_5$. The combination of temperature-dependent X-ray absorption and first-principles studies clearly indicate the Inverse Star of David structure as the preferred reconstruction in the low-temperature CDW phase. Our results highlight the critical role that Sb orbitals plays and establish orbital hybridization as the direct mediator of the CDW states and structural transition dynamics in Kagome unconventional superconductors. This is a significant step towards the fundamental understanding and control of the emerging correlated phases from the Kagome lattice through the orbital interactions and provide promising approaches to novel regimes in unconventional orders and topology.

# 1. Introduction

Spontaneous symmetry breaking is a quintessential concept in condensed matter physics. Landau's theory of phase transition emphasizes the importance of symmetry and links phase transitions to symmetry-breaking phenomena such as superconductivity,[1, 2] superfluidity and Bose-Einstein condensation,[3, 4] long-range charge-density wave (CDW)[5, 6] and magnetic transition.[7] Symmetry-breaking phenomena are often found in quantum systems where the complex interplay of charge, spin, lattice, and orbital degrees of freedom takes place.[8, 9] Among other macroscopic quantum phenomena, the periodic distortion of the long-range CDW is well recognized for the role it plays at the atomistic level. Yet the mechanism that underlies its competitive or complementary relationship with unconventional superconductivity[10, 11] remains unresolved. Recent reports of exotic chiral orders in the layered Kagome Dirac metals, $AV_3Sb_5$ (A = K, Rb or Cs),[12] are tantalizingly appealing to the research community. With their unique lattice geometry, Kagome Dirac metals host a diverse range of exotic properties and topologically-nontrivial states, exhibiting giant anomalous Hall conductivity,[13] magneto-quantum oscillations,[14] topological charge order,[12, 15] and superconductivity.[16, 17] Although there are reports of normal-state CDW of the topological chiral charge order [12] and diverse superconducting gaps structures in Kagome materials,[18] the roles of fundamental properties such as Fermi surface nesting,[19] electron-phonon coupling,[20] and orbital hybridization play in these symmetry-breaking phenomena remain largely unclear. Notably, this interest is further catalyzed by the effects of hybridization between different constituent atomic orbitals and their close association with various quantum phase transition processes.[21-25]

Here we report the direct participation of the V$3d$—Sb$5p$ orbital hybridization in mediating the CDW transition dynamics in CsV$_3$Sb$_5$. The combination of temperature-dependent X-ray absorption spectroscopy and first-principles studies in this comprehensive study further reveals the Inverse Star of David (ISD) as the preferred structure in the CDW phase. Contrary to conventional view where long-range charge order is mediated solely by the vanadium atoms,[20, 26, 27] this study unambiguously highlights the pivotal role that the constituent antimony orbitals play on the formation of van Hove singularity structures and the stability of the CDW states in AV$_3$Sb$_5$ systems, through their extensive interaction and the complementary effects between the V- and Sb-atoms on the electronic structures (**Figure 1**a).[28-30] Our study additionally gains importance in light of the recent report where Sb–oxygen covalency contributes to the emergence of superconductivity in antimonates.[31] Unraveling the mechanism that governs the CDW states in CsV$_3$Sb$_5$ Kagome systems provides essential hints to identify the key ingredients of Kagome unconventional superconductors and further serves as a platform to uncover the interplay between the symmetry-breaking orders[19, 32] and other unconventional orders and topologies.

## 2. Results and Discussion

### 2.1. Validation of Sample Quality

Layered CsV$_3$Sb$_5$ single crystals were synthesized via a self-flux growth method.[16] This class of AV$_3$Sb$_5$ materials exists in both the conventional 1×1 hexagonal crystal structure,[13] and the √3×√3R30° reconstruction.[32] Figure 1b displays the Scanning tunneling microscopy (STM) topographic image of the Cs-terminated surface which indicates the typical single-unit-cell terrace of ~ 9.4 Å (Figure 1c).[13, 33] The two unique cleaved Cs surface morphologies along the red dotted lines (Figure 1d) are plotted in Figure 1e where they correspond to the 1×1 hexagonal and √3×√3$R$30° structures, respectively.

The very high crystallinity and quality of the $CsV_3Sb_5$ sample is further confirmed via high-resolution X-ray Diffraction (HR-XRD), in its diffraction pattern and its regular hexagonal structure (**Figure 2**a).[34] Reciprocal space mappings (RSMs) along the (004)$_{HL}$ and ($\bar{1}$05)$_{HL}$ orientations (Figure 2b and c, respectively) further confirmed the good crystalline property of the single-crystal sample via the weakly diffusing features B and C from the main peak A.

The presence of the CDW states in the $CsV_3Sb_5$ sample is confirmed through the magnetic susceptibility, $\chi(T)$, where it exhibits a sharp decline at ~94 K for $\mu_0 H$=1 T and $H//ab$ (Figure 2d).[13, 16] Temperature-dependent magnetic susceptibility $4\pi\chi(T)$ under zero-field-cooling (ZFC) and field-cooling (FC) modes at 1 mT for $H//ab$ (Figure 2e) also confirms the superconductive state in this sample at $T_C$~2.6 K.[16]

## 2.2. Temperature-dependent X-ray Absorption Spectroscopy Characterization

To investigate the evolving electronic structures and orbital-coupling properties of $CsV_3Sb_5$ in temperature ranges near $T_{CDW}$, the temperature-dependent X-ray absorption spectroscopy (XAS) is conducted to examine how the V3$d$ orbital evolve in this temperature range. The characteristic peaks registered by the XAS measurements correspond to the unoccupied band above the Fermi surface, which is strongly sensitive to lattice symmetry, crystal field splitting and orbital hybridization.[35, 36] **Figure 3**a displays the temperature-dependent V L-edge XAS spectra of the $CsV_3Sb_5$ sample over a wide temperature range around $T_{CDW}$. In the XAS spectra, where shoulders $A^*$ (~517.0 eV), $B^*$ (~518.2 eV) and the characteristic peaks $C^*$ and $D^*$ (~519.2 and ~525.6 eV, respectively) are observed. Feature $A^*$ is attributed to the slight hybridization between atomic orbitals while $C^*$ and $D^*$ are the L$_{2,3}$-edges representing the V2$p_{3/2}/p_{1/2}$→3$d$ electronic transitions,[37] all displaying very weak temperature variation that falls within the experimental error range between 40 and 300 K. While feature $C^*$ registers a slight temperature variation, there is no clear temperature-dependent trend that is noticeable. Conversely, we noticed a prominent shoulder $B^*$ (denoted by black solid arrow) before the L$_3$-edge displaying very strong and clear temperature-dependent unlike the other aforementioned features in the XAS spectra. To better distinguish the temperature-dependent intensity of the respective features, an intensity differential map, $\Delta\mu=\mu(T)-\mu(T_{CDW}$=94 K), is plotted for the entire temperature range between 40 and 300 K (Figure 3b) with the XAS spectrum at $T_{CDW}$=94 K taken as reference. The intensity differential, $\Delta\mu$, displays very strong temperature-dependent fluctuation at ~518.2 eV where feature $B^*$ is located, while remaining largely unchanged in the other spectral regions, especially where the V-L$_{2,3}$ absorption edges are located. These are indications that the V3$d$ orbital do not mediate or participate in the formation of the CDW states alone. Instead, the strong temperature-dependence of shoulder $B^*$ particularly near $T_{CDW}$ provides strong suggestion to investigate in detail the interaction and hybridization of the V3$d$ orbital with their neighbouring electronic bands.[38] Notably, the roles of orbital hybridization in mediating the CDW phase transition are not merely restricted to Kagome systems. Instead, the stability and formation of CDW states have also been reported to be mediated by such hybridization effects in multiple systems including two-dimensional transition-metal dichalcogenides[39] and unconventional superconductors,[40] where they instigate the energy band reconstruction that enables the gap opening. This in turn facilitates the formation of the CDW states. Hence, the combination of our study and previous reports attests to the complementary effects of the V- and Sb-atoms in the form of V$3d$—Sb$5p$ orbital hybridization in dictating the electronic structures and CDW states in AV$_3$Sb$_5$ systems.[28-30] As we will show later by a combination of temperature-dependent analysis of feature $B^*$ and a series of first-principles studies, it can be deduced that V$3d$—Sb$5p$ orbital hybridization is directly involved in the formation of the CDW states.

Detailed temperature-dependent analysis of feature $B^*$ in the XAS spectra is performed. Figure 3c displays the trend with decreasing temperature from 300K, where the peak intensity first decreases with decreasing temperature and eventually sharply minimizes at $T_{CDW}$=94 K. Below 94 K, the intensity of $B^*$ rises again. A detailed analysis of the intensity evolution of feature $B^*$ is given by segmenting the temperature range into four main temperature regions: Region I (~140-300 K), Region II just above $T_{CDW}$ (~94-140K), Region III just below $T_{CDW}$ (~70—94 K), and Region IV at lowest temperatures (30-~70 K).

At high-temperature Region I, the Kagome structure is dominated by strong thermal vibrations where effects such as the inter-orbital hybridization and electron-electron correlations are overshadowed by phonon interactions. Unique thermal behavior in the form of lattice vibrations has been reported to take place in Kagome materials in this temperature region, which significantly overshadow the orbital hybridization effects.[30, 41, 42] Conversely, thermal effects and phonon interactions are suppressed at low temperature (in temperature Region IV below ~70 K). In this temperature region, the intensity of feature $B^*$ rises continuously to approximately back to its original intensity as temperature decreases to 40 K. The effects observed in temperature Region IV can be attributed to the strong electron-electron correlations associated with another nematic symmetry-breaking order[5, 19] dominating in this low-temperature range, which screens the effects of orbital hybridization.

Nevertheless, as we consider temperature Regions II and III that is located close to the CDW transition temperature ($T_{CDW}$=94 K), with the temperature decreasing and as it approaches $T_{CDW}$ in temperature Region II, feature $B^*$ registers an abrupt intensity drop (approximately gradient: -0.065) and minimizes at $T_{CDW}$ =94 K. Below 94 K in temperature Region III, the intensity of feature $B^*$ then exhibits a rapid upturn with a gradient of ~0.050 (see inset of Figure 3c). The unique temperature trends in regions II and III near $T_{CDW}$ provide important suggestions of electronic orbital behaviors in mediating the CDW transition.

## 2.3. Density of States Analysis

To provide further insights into the orbital behaviors between the pristine and CDW states and to better track the transient properties taking place at $T_{CDW}$, detailed first-principles calculations have been conducted. While the Kagome lattice is known to be in its conventional pristine structure above $T_{CDW}$, CDW transition is accompanied by a three-dimensional 2×2×2 lattice reconstruction associated with the movement of the constituent V atoms.[12, 19] Previous studies have shown that the 'Star of David'(SOD) and 'Inverse Star of David' (ISD) structures are possible resulting structures below $T_{CDW}$, as they exhibit enhanced structural stability compared to their high-temperature pristine counterpart.[19] Nevertheless, there is no consensus on a preferred structure or if they co-exist in the CDW state.[26, 43]

Figure 3d presents the contributions from the respective atomic orbitals to the density of states (DOS) of the $CsV_3Sb_5$ Kagome lattice in its high-temperature pristine state. As expected, the V3$d$ orbital has dominant contribution to the DOS near the Fermi level. Moreover, the Sb5$p$ orbital also plays a pivotal role as evidenced by a much larger contribution to the DOS than the other constituent orbitals. The significant DOS contributions by the V3$d$ and Sb5$p$ orbitals also apply to the low-temperature ISD and SOD structures (see Supplemental Material Figure S3). The domination of the two main electronic states, V3$d$ and Sb5$p$ orbitals, near the Fermi level are therefore clear indications that there is a presence of strong V3$d$—Sb5$p$ orbital hybridization taking place.

To further confirm the central role that V3$d$—Sb5$p$ hybridization plays during the CDW transition, the PDOS of the V3$d$ and Sb5$p$ orbitals (**Figure 4**a and b, respectively) are compared in the pristine, SOD and ISD states. While the PDOS in the pristine and SOD states are almost identical, we note the significant difference in the ISD structure especially near feature σ. The variation of the PDOS and the changes in energy positions of feature σ in both the V3$d$ and Sb5$p$ orbitals have a direct effect on the hybridization strength. While feature $σ$ of both orbitals shifts to higher energies, its intensity in the V3$d$ orbital remains unchanged, and that of Sb5$p$ is enhanced (Figure 4b). In terms of feature $α$, while the V3$d$ orbital does not show any changes in PDOS between different structures (Figure 4a), the PDOS component of the Sb5$p$ orbital in the 0.8—1.2 eV energy range (Figure 4b) displays a very slight but noticeable difference between the pristine and ISD structures. Whereas the $α$ feature of the SOD structure is practically identical with its pristine counterpart. While these differences may be minor, the variation of in the PDOS component of the Sb5$p$ orbital may play a role in modifying the V3$d$—Sb5$p$ hybridization between different structural phases. Meanwhile, the larger contribution by the Sb5$p$ orbital in terms of the $σ$ feature in the PDOS strengthens the V3$d$—Sb5$p$ hybridization in the ISD structure.

Note that the PDOS of the V3$d$ orbital (Figure 4a) near the Fermi level yields identical features that are experimentally derived from the V L-edge XAS spectrum near 519 eV (Figure 3a). Features $α$ and $σ$ of the V3$d$ orbital PDOS (Figure 4a) are responsible for V L-edge features $C^*$ and $B^*$, respectively (Figure 3a). The energy separation of ~0.8 eV between features $σ$ and $α$ in the PDOS calculation is also consistent with the separation between features $B^*$ and $C^*$ in the V L-edge spectra. It is safe to conclude that the temperature-dependent behavior of feature $B^*$ is attributed to the change in feature $σ$ of the PDOS during the CDW transition. Since feature $σ$ of both the V3$d$ and Sb5$p$ orbitals changed under ISD lattice distortion but remains virtually unchanged in the SOD state, the temperature-dependent behavior in the XAS spectra therefore indicates pristine-to-ISD lattice at $T_{CDW}$.

To further investigate how the V3$d$—Sb5$p$ hybridization evolves during the CDW transition, PDOS of both the V3$d$ and Sb5$p$ orbitals are calculated separately as functions of lattice distortion between the pristine (initial) and ISD (final) states (see Supplemental Material for the simulation of the pristine-to-SOD transition process). Since the lattice distortion process can be regarded as the straight motion of V-atoms due to space group and symmetric properties,[19] the fractional distortion can therefore be defined as $\gamma = d/d_0$, where $d$ represents the V-atom displacement while $d_0$ denotes the complete displacement of the V-atom when the final ISD structure is obtained. Figure 4c and d display the PDOS of V3$d$ and Sb5$p$ orbitals, respectively, at the intermediate stages of the pristine-to-ISD transition, depicting how $\sigma$ evolves due to its association with the temperature-dependent feature $B^*$ in the XAS spectra. While the intensity of $\sigma$ changes non-trivially with $\gamma$, $\sigma$ of both the V3$d$ and Sb5$p$ orbitals gradually shifts to a higher energy with increasing $\gamma$ distortion.

## 2.4. Tracking the Pristine-to-ISD Structural Transition Process

To better track the evolution of feature $\sigma$ as a function of $\gamma$, Figure 4e and f plot its relative intensity and position, respectively, in its intermediate states. For lattice distortion $\gamma \lessapprox 0.6$, the intensity of $\sigma$ belonging to the V3$d$ orbital is suppressed with increasing $\gamma$ (Figure 4e). Meanwhile, the intensity of σ belonging to the Sb5$p$ orbital remains nearly unchanged. This gives rise to a weaker V3$d$—Sb5$p$ orbital hybridization strength due to the suppression in the DOS contribution by the V3$d$ orbital. This explains the significant drop in the intensity of feature $B^*$ of the XAS spectrum as $T$ approaches $T_{CDW}$ in temperature Region II (Figure 3c). With the progressing lattice distortion at $\gamma>0.6$, the intensity of $\sigma$ for both the V3$d$ and Sb5$p$ orbitals are enhanced along with a significant drop in terms of their energy difference (Figure 4f). This leads to an enhancement in the V3$d$—Sb5$p$ orbital hybridization, which in turn brings about a significant increase in the intensity of feature $B^*$, as reflected in the XAS spectrum in temperature Region III (Figure 3c). We can therefore attribute the temperature-dependent behavior of feature $B^*$ in the XAS spectrum to the changes in the V3$d$—Sb5$p$ orbital hybridization during the CDW transition process, during which the pristine-to-ISD transition can be regarded as a continuous lattice deformation process. As discussed in Figure S7 of the Supplemental Material, pristine-to-SOD transition is ruled out during the CDW transition.

## 3. Conclusion

Our work points to the critical role that hybridization between the V and Sb orbitals plays in mediating the formation of the CDW states in the CsV$_3$Sb$_5$ Kagome unconventional superconductor. Demonstrated direct participation of the Sb atomic orbitals in the CDW transition process opens new discussions on the long-held view that the long-range charge order is mediated solely by vanadium atomic orbitals. This finding is also relevant to general understanding of the critical roles that Sb orbitals play in electronic phase transitions in different materials, bearing in mind the recent report of their direct involvement in the interplay between superconductivity and CDW orders in antimonates. Apart from providing convincing evidence to the pristine-to-ISD structural transition at $T_{CDW}$, our study also provides further insights to the lattice distortion during the intermediate stages of this CDW transition process where the structural transformation is shown to behave in a manner that resembles thermal molecular motion. Ultimately, our study delivers a significant breakthrough towards the fundamental understanding and control of the correlated phases emerging from the Kagome lattice. Not only does it renew previous discussions on the roles that constituent atomic orbitals play in the CDW and superconductive properties in Kagome materials,[44] but it also provides new exploratory insights towards orbital origins of other novel topological and unconventional orders.

## 4. Methods Section

**Sample Synthesis:** $CsV_3Sb_5$ single crystals were grown by spontaneous nucleation in a self-flux method. Firstly, a mixture of high-purity Cs ingots, V and Sb powders (with a molar ratio of 10:3:28, and total mass of about 3g) was loaded in an alumina crucible. The alumina crucible was jacketed in a Ta tube. Thereafter, the Ta tube was welded through Argon arc melting and then sealed in an evacuated quartz ampoule. Subsequently, the quartz ampoule was loaded into a muffle furnace, heated up to 1193 K, and maintained at this temperature for 24 hours. The furnace was then gradually cooled to 773 K at a rate of 2 K/h. The obtained crystals were wrapped by flux which could then be washed off by alcohol. Shiny millimeter-size plate-like crystals were then harvested. The crystals are stable in air for several weeks.

**Scanning Tunneling Microscopy Measurements:** The samples are exfoliated in-situ at room temperature (300 K) in a preparation chamber, high-resolution STM measurements were then performed at 77 K at ultrahigh vacuum condition ($\sim 1 \times 10^{-10}$ mbar) in an Omicron UHV system interfaced to a Nanonis controller. Electrochemically etched tungsten tips were used as the grounded probe while the sample holder has a voltage bias applied to it.

**High-resolution X-ray Diffraction and Reciprocal Space Mapping Measurements:** High-resolution X-ray diffraction (HR-XRD) experiments are performed at the X-ray Demonstration and Development (XDD) beamline at the Singapore Synchrotron Light Source (SSLS).

**Magnetization measurements:** Temperature dependence of magnetic susceptibility was measured on Quantum Design Magnetic Property Measurement System (MPMS3).

**X-ray Absorption Spectroscopy:** Temperature-dependent X-ray absorption spectroscopy (XAS) measurements were performed on freshly cleaved $CsV_3Sb_5$ single crystal samples at the Soft X-ray–ultraviolet (SUV) beamline at the Singapore Synchrotron Light Source, in a vacuum chamber with a base pressure of $\sim 1 \times 10^{-9}$ mbar via the Total Electron Yield (TEY) mode.[45, 46] The incident X-ray is projected at a normal incident angle onto the sample surfaces. The XAS measurements were in a descending temperature sequence (i.e., 300 K → 200 K → 100 K → 98 K → 96 K → 94 K → 92 K → 90 K → 40 K). About 20 mins of waiting time is given at each temperature point before the start of each measurement.

**First-Principles Calculations:** First-principles calculations were performed using the Vienna ab initio simulation package (VASP) code,[47, 48] implementing density functional theory (DFT). For the electron exchange-correlation functional, we used the generalized gradient approximation (GGA) in the form proposed by Perdew, Burke, and Ernzerhof (PBE).[49] The energy cutoff of the plane-wave basis was set to 400 eV and the zero damping DFT-D3 van der Waals correction was employed throughout the calculations.[50] In the pristine phase, the atomic positions and lattice vectors were fully optimized using the conjugate gradient (CG) scheme until the maximum force on each atom was below 0.01 eV/Å, with an energy precision of $10^{-5}$ eV. The Brillouin zone (BZ) was sampled using a 19×19×11 Γ-centered Monkhorst-Pack grid. For the 2×2×1 phase, the energy precision was set to $10^{-7}$ eV and a *k*-mesh of 9×9×11 (19×19×11) was used in stationary (density of state, DOS) calculations.[51]

**Acknowledgements:**
This work was supported in part by the Strategic Priority Research Program of the Chinese Academy of Sciences, Grant No. XDB25000000, National Natural Science Foundation (52172271), the National Key R&D Program of China No. 2022YFE03150200. J.W. acknowledges the Advanced Manufacturing and Engineering Young Individual Research Grant (AME YIRG Grant No.: A2084c170). C.S.T., P.Y. and C.D. are supported by SSLS via NUS Core Support C-380-003-003-001. M.V.M. acknowledges support from the Research Foundation-Flanders (FWO-Vlaanderen). The authors acknowledge the Singapore Synchrotron Light Source for providing the facility necessary for conducting the research. The Laboratory is a National Research Infrastructure under the National Research Foundation, Singapore. Any opinions, findings and conclusions or recommendations expressed in this material are those of the author(s) and do not reflect the views of National Research Foundation, Singapore.

**Conflict of interest**
The authors declare no conflict of interest.

**Author Contributions:**
X.Y. conceived the project. L.Y. synthesized the samples with help from C.G. S.H. and C.S.T. performed the XRD and XAS experiments and analysed the data with help from P.Y., C.D. and X.Y. J.G. performed the STM experiments and analysed the data. L.L. performed the DFT calculations. L.L, L.Z., M.Z. and M.V.M. contributed the theoretical interpretations. S.H. and C.S.T. wrote the manuscript, with input from all the authors. S.H., C.S.T., L.L. and Y.L contributed equally to this work.

# Figures

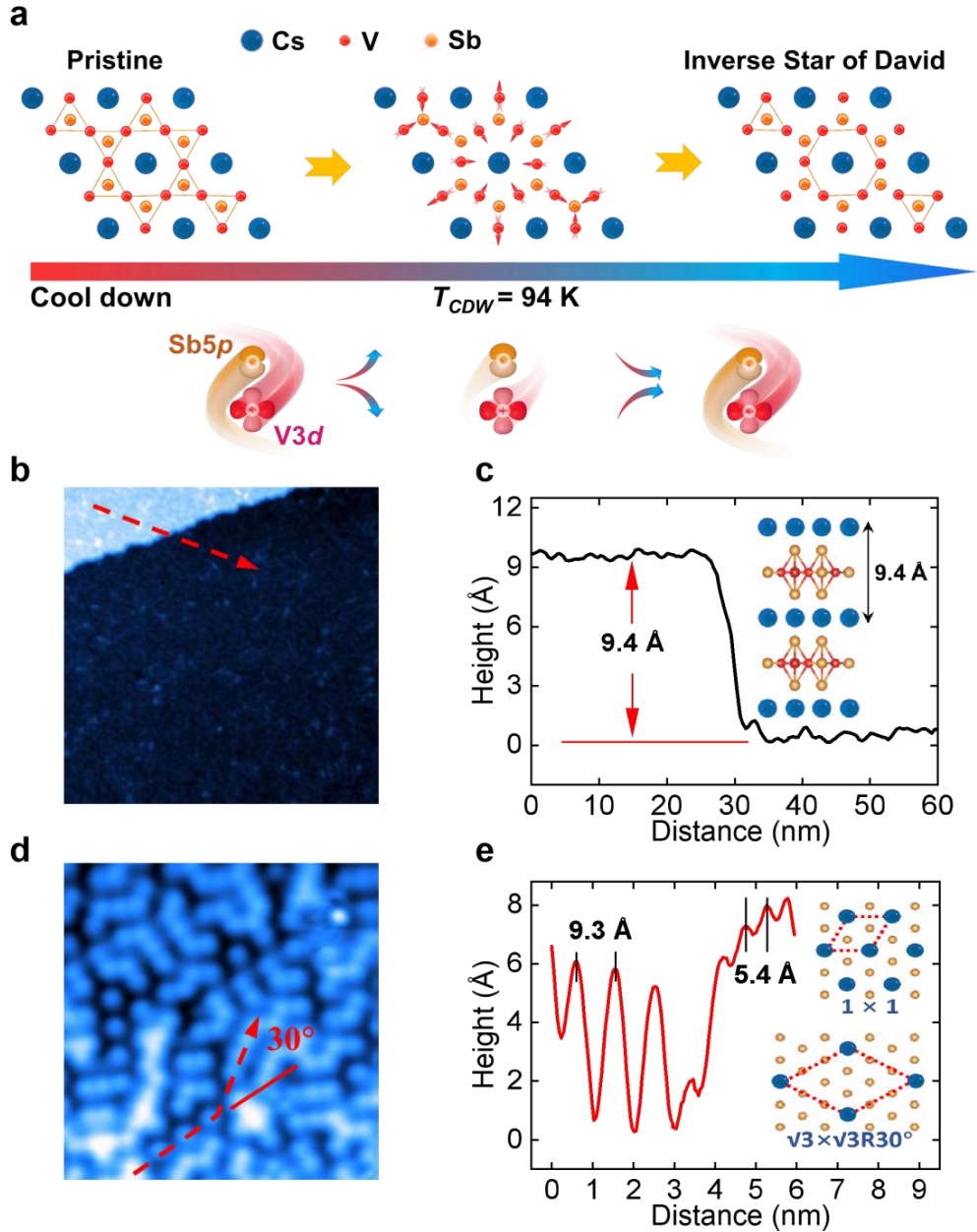

**Figure 1.** Structural transition dynamics at $T_{CDW}$ and morphology of the CsV$_3$Sb$_5$ single crystal samples. a) Schematic depiction of the changes in V3$d$—Sb5$p$ orbital hybridization strength during the structural transition of CsV$_3$Sb$_5$ from pristine to Inverse Star of David (ISD) structure at $T_{CDW}$. b) Topographic image (setpoint: $V$ = -0.1 V, $I$ = 30 pA) of the Cs-terminated terrace to the bottom Sb surface. c) Line profile along the red dashed line in (b). Inset represents the side view of the atomic structure in CsV$_3$Sb$_5$. d) High-resolution STM image of the two main types of cleaved Cs surface morphologies: $\sqrt{3}\times\sqrt{3}R30°$ and 1×1 (setpoint: $V$ = 0.1 V, $I$ = 50 pA). e) Line profile along the red dashed line in (d) shows the measured lattice period of ~9.3 Å belonging to the $\sqrt{3}\times\sqrt{3}R30°$ configuration and the period ~5.4 Å of the conventional 1×1 hexagonal structure. Inset shows the atom arrangement of the respective surface structures – $\sqrt{3}\times\sqrt{3}R30°$ and 1×1.

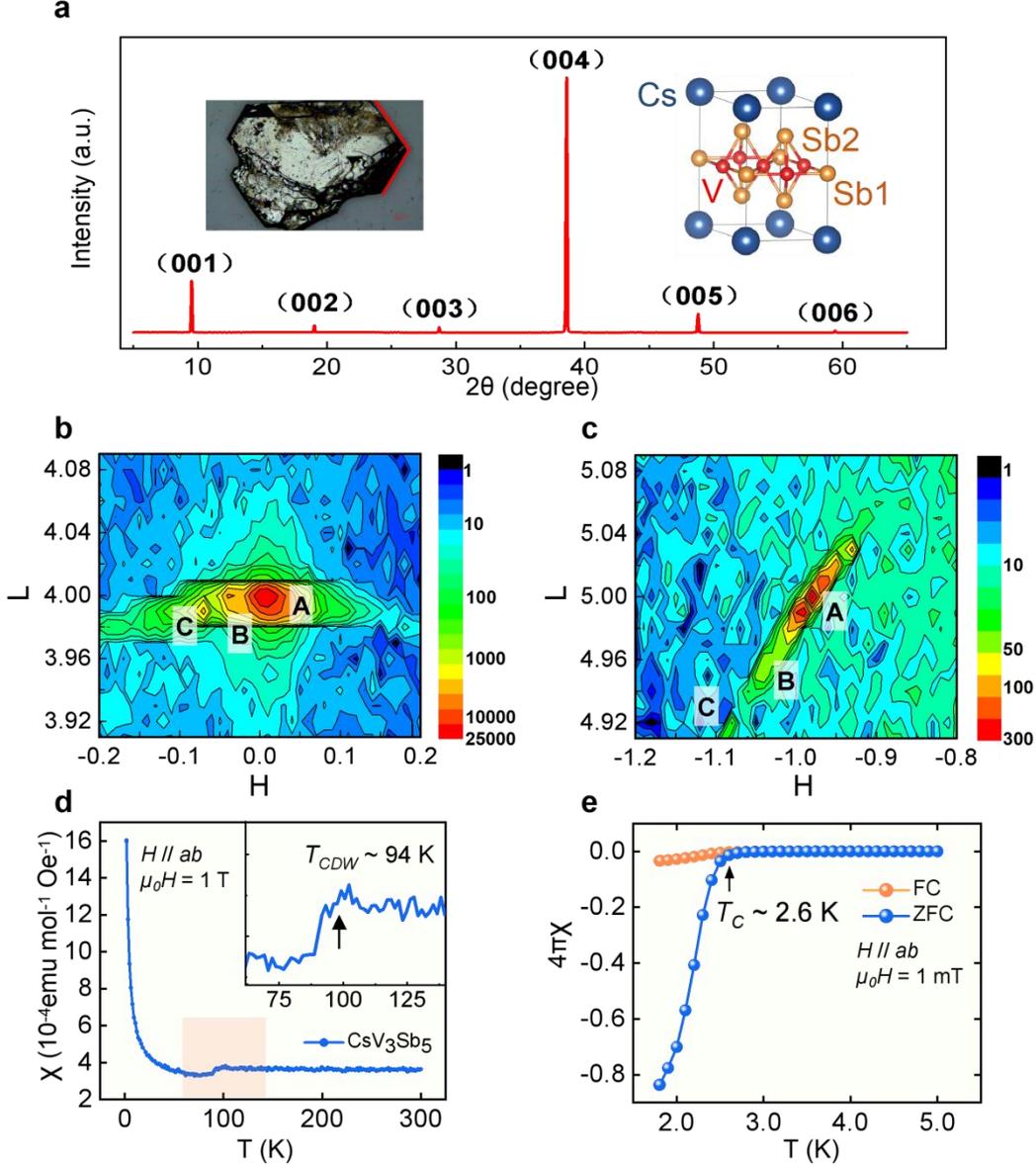

**Figure 2.** Structural and magnetic susceptibility properties of CsV$_3$Sb$_5$. a) High-resolution X-ray Diffraction (HR-XRD) pattern. Insets: the optical image of the sample and the unit structure of CsV$_3$Sb$_5$. b) Reciprocal space mapping (RSM) in the (004)$_{HL}$ and c) ($\bar{1}$05)$_{HL}$ reciprocal space region of the sample. d) Magnetic susceptibility χ(T) at $\mu_0H$ = 1 T for *H//ab*. Inset: Zoomed-out χ(T) curve of the shaded temperature range indicating $T_{CDW}$ transition. e) Magnetic susceptibility, 4πχ(T), with ZFC and FC modes at $\mu_0H$ = 1 mT for *H//ab*, results indicate the onset of superconductive transition at $T_C$~2.6 K.

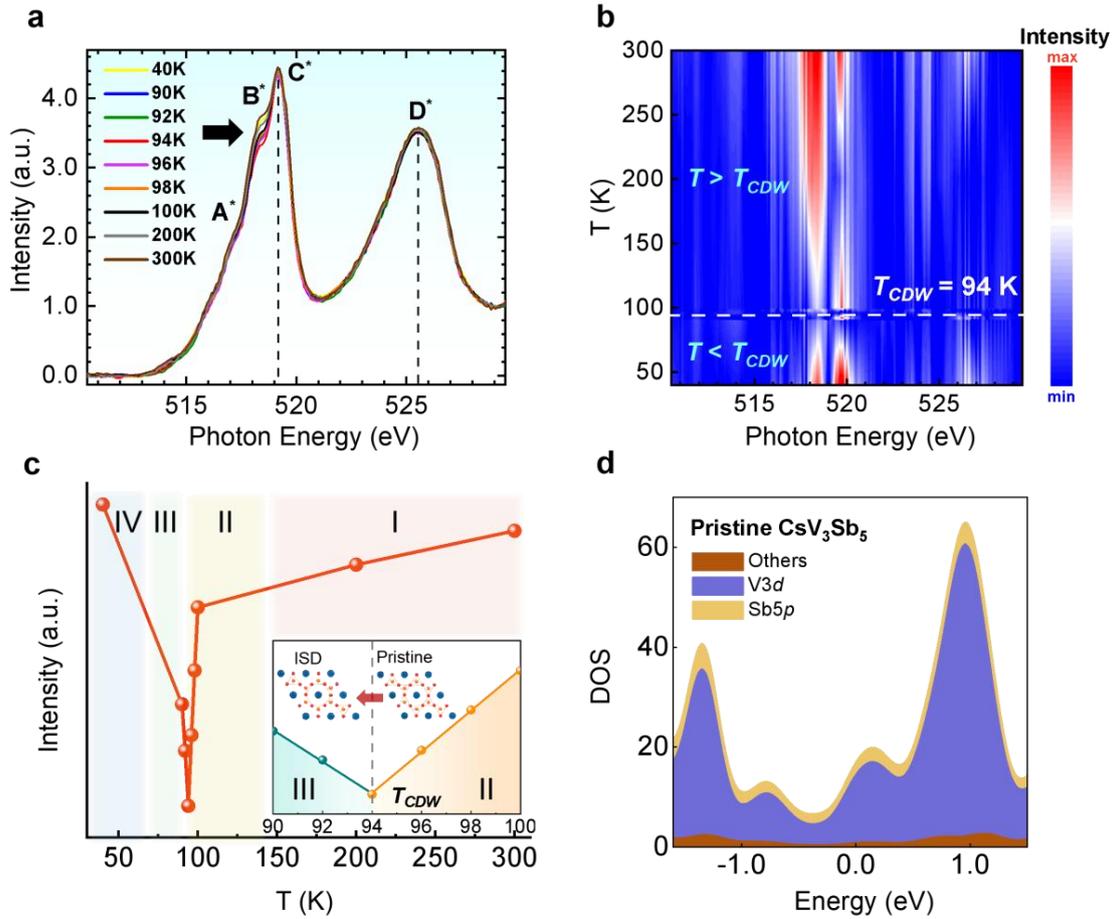

**Figure 3.** Temperature-dependent XAS characterization, Density of States (DOS) and Partial Density of States (PDOS) analyses for different lattice structures of $CsV_3Sb_5$. a) Temperature-dependent V $L_{2,3}$-edge XAS spectra, and b) Differential XAS spectral intensity ($\Delta\mu=|\mu(T)-\mu(T=94\ K)|$) derived from the XAS spectra. c) Temperature-dependent intensity of feature $B^*$ in four separate temperature regions between 300 K and 40 K. Inset: Intensity variation of feature $B^*$ in temperature regions II and III located near $T_{CDW}$. d) DOS of pristine state $CsV_3Sb_5$ near the Fermi level with PDOS contributions from the atomic orbital constituents of V$3d$, Sb$5p$ and other orbitals (see Supplemental Material for DOS of SOD and ISD structures).

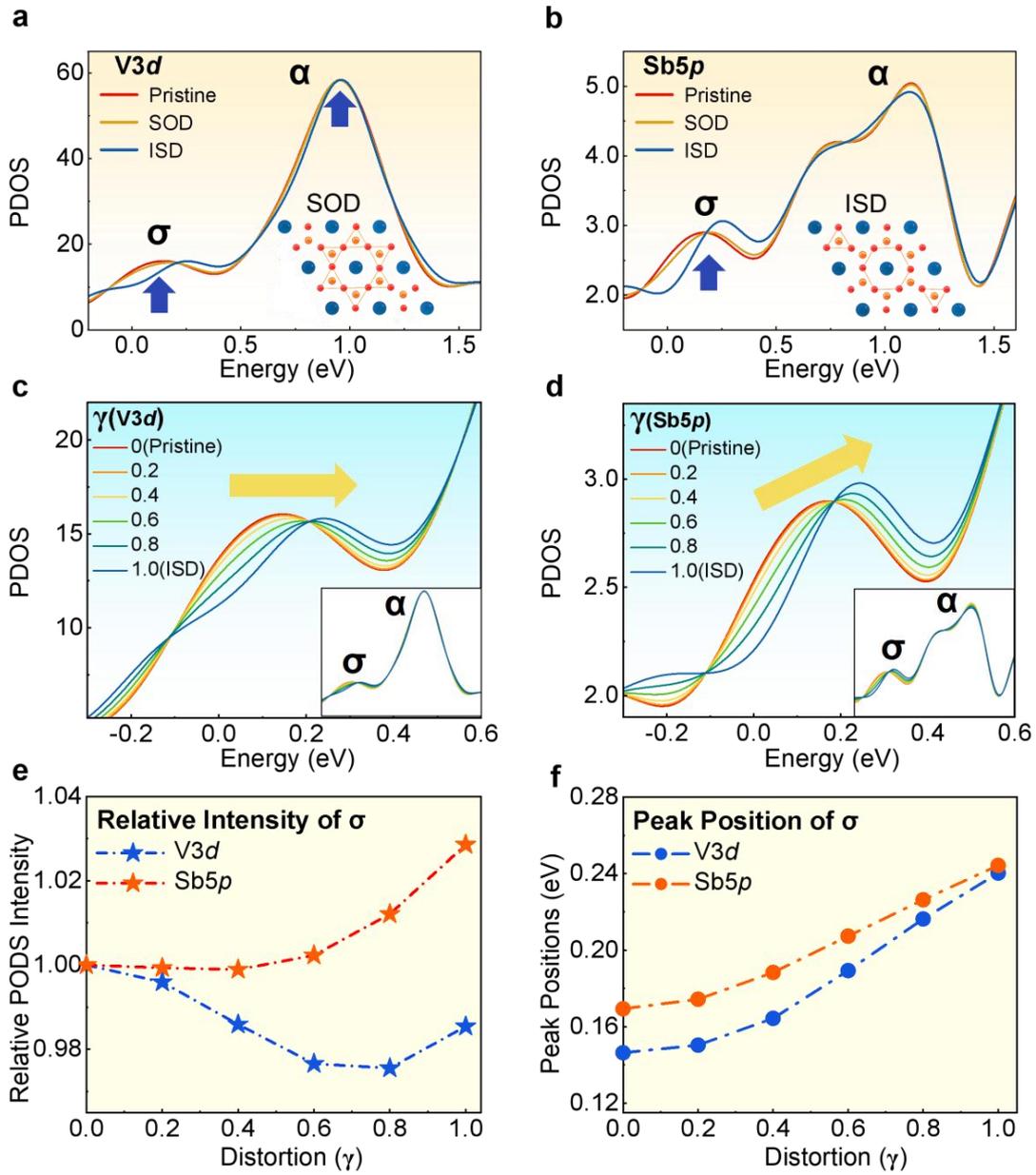

**Figure 4.** Partial Density of States (PDOS) of the respective constituent orbitals for different lattice structures of $CsV_3Sb_5$. a) PDOS of the V3$d$ orbital. Inset: Schematic of the SOD lattice. b) PDOS of the Sb5$p$ orbital. Inset: Schematic of the ISD lattice. c) Changes in the PDOS of the constituent V3$d$, and d) Sb5$p$ orbitals, as functions of ISD lattice distortion, $\gamma$, during the intermediate steps of the pristine-to-ISD transition. Changing PDOS profile of feature $\sigma$ with increasing $\gamma$ has been tracked and indicated by yellow arrows. Insets: Zoomed-out PDOS of the V3$d$ and Sb5$p$ orbitals, respectively. e) Changes in the relative intensity, and f) position of feature $\sigma$ for the PDOS of both V3$d$ and Sb5$p$ orbitals during the intermediate steps of the pristine-to-ISD transition.